\documentclass{aa}  
\usepackage{natbib}
\usepackage{float} 
\bibpunct{(}{)}{;}{a}{}{,} % to follow the A&A style
\usepackage{graphicx}
\usepackage{txfonts}
\usepackage{lipsum}
\usepackage{subcaption}         % necessary for continued figures, example in section 3
\usepackage{xcolor} % for \textcolor
\usepackage{lscape}             % to rotate a single page table, example in appendix.
\usepackage{placeins}           % useful with \FloatBarrier, to keep 
\usepackage{hyperref}
\usepackage{siunitx}
\usepackage{caption}
\usepackage{subcaption}
\usepackage[dvipsnames]{xcolor}
\usepackage{tabularx}
\usepackage{booktabs}
\usepackage{makecell}
\hypersetup{
    colorlinks=true,
    linkcolor=blue,
    urlcolor=blue,
    citecolor=blue
    }
\RequirePackage{etex} 
\begin{document}

   \title{Collisional excitation of H$_2$CO by He}

   \subtitle{Experimental validation of state-of-the-art scattering calculations}

%%%%%%%%%%%%%%%%%%%%%%%%%%%%%%%%%%%%%%%%
% Please do not include ORCIDs next to author names.
% Only ORCIDs authenticated by individual authors in EDP Sciences editorial system will be taken into account.
% ORCIDs included here will be removed.
%%%%%%%%%%%%%%%%%%%%%%%%%%%%%%%%%%%%%%%%
   \author{Chinmai Sai Jureddy\inst{1}
        \and Sandor Demes\inst{2}
        \and Francesca Tonolo \inst{1}        
        \and Francois Lique\inst{1}\fnmsep\inst{3}\fnmsep\thanks{Email: francois.lique@univ-rennes.fr}
        \and Ian R. Sims\inst{1}\fnmsep\inst{3}\fnmsep\thanks{Email: ian.sims@univ-rennes.fr}
        }

   \institute{Univ Rennes, CNRS, IPR (Institut de Physique de Rennes) - UMR 6251, F-35000 Rennes, France
   \and HUN-REN Institute for Nuclear Research (ATOMKI), Debrecen, Hungary
   \and Institut universitaire de France (IUF)}

   \date{Received June 1, 2026}

  \abstract
   {Non-local thermodynamic equilibrium conditions in the interstellar medium require collisional rate coefficients to model astronomical observations; these are usually determined from theoretical scattering calculations.}
   {The aim of this study is to measure experimentally low-temperature pressure-broadening cross-sections for the H$_2$CO-He system in order to validate the theoretical methodology involved in determining new collisional rate coefficients.} 
   {The experiments employed the chirped-pulse in uniform supersonic flow method, and H$_2$CO is generated \textit{in situ } by 193 nm excimer laser photolysis of tetrahydrofuran in cold He flows. State-of-the-art calculations are performed by computing a new potential energy surface for the H$_2$CO–He system which is subsequently implemented in scattering calculations using the close-coupling method to derive pressure broadening cross-sections and collisional rate coefficients.}
   {Excellent agreement between theory and experiment is obtained, with the calculated values falling within the 95\% confidence intervals of the experimental measurements. Such agreement validates the high accuracy of the theoretical data.}
   {Helium constitutes about 20\% relative to H$_2$ in the interstellar medium. The inclusion of collisional rate coefficients for H$_2$CO with He in radiative transfer modelling leads to variations in the excitation temperature of frequently detected rotational lines of up to 12\% in warm regions such as protostars.}


   \maketitle
   \nolinenumbers

%%%%%%%%%%%%%%%%%%%%%%%%%%%%%%%%%%%%%%%%%%%%%%%%%%%%%%%%%%%%%%
\section{Introduction}
Formaldehyde (H$_2$CO) has been widely detected in the interstellar medium (ISM) since its first identification by \citet{snyder_microwave_1969} and early observations already revealed the effects of non-local thermodynamic equilibrium (non-LTE) conditions. For example, anomalous absorption of the 1$_{1,0}$$\leftarrow$1$_{1,1}$ rotational transition of H$_2$CO was found, implying level populations corresponding to a Boltzmann distribution at a temperature of approximately 1.8~K, lower than the 2.73~K cosmic microwave background radiation \citep{palmer_formaldehyde_1969}, and much lower than in typical cold molecular clouds ($\sim10$~K). This cooling effect (often called "anti-maser" or maser action) in H$_2$CO can only be explained by a collisional pumping mechanism \citep{evans1975interstellar}. This was first interpreted using classical treatments of inelastic collisional propensity rules \citep{townes1969pumping} and was subsequently validated by quantum mechanical calculations \citep{garrison1975cooling}. The excess population in the 1$_{1,1}$ state is due to favorable collisional excitation to the 2$_{1,2}$ state over the 2$_{1,1}$ state and its subsequent radiative decay overpopulate the 1$_{1,1}$ state \citep{townes1969pumping}. Such masers are also observed in the 2$_{1,1}$$\leftarrow$2$_{12}$ transition and the 1$_{1,1}$$\leftarrow$1$_{1,0}$ transition of H$_2$$^{13}$C$^{16}$O \citep{evans1975interstellar}. This illustrates the prevalence of non-LTE conditions in astrophysical media, where both radiative and collisional processes must be accounted for to determine molecular level populations. 

Transitions observed under non-LTE conditions can serve as robust diagnostic tools for constraining physical conditions, such as temperature and density, in interstellar space. Due to their widespread presence, H$_2$CO transitions are used in astronomy as tracers of many astrophysical environments \citep{muhle2007formaldehyde,mahmut_formaldehyde_2024,christensen_formaldehyde_2026} and to constrain the ortho-to-para ratio of  H$_2$ \citep{troscompt_constraining_2009}.

The use of rotational transitions observed under non-LTE conditions requires accurate collisional (de-)excitation rate coefficients for  the observed molecular species with the dominant interstellar collision partners, namely H, H$_2$, He, and electrons \citep{roueff_molecular_2013}. These rate coefficients are obtained entirely from theoretical calculations and therefore rely on both the accuracy of the underlying potential energy surface (PES) and the quantum scattering methodology used to generate the rate coefficients. However, experimental validation of these calculations is very rare, especially at temperatures (and collision energies) of relevance to the interstellar medium (ISM), and in particular interstellar molecular clouds (ISCs) where the temperature can range from 10~K or even below in dense ISCs or cold cores to $\sim100$~K for diffuse clouds.  Experimental validation is  essential to assess the uncertainties in these theoretical quantities and, by extension, in the physical conditions inferred from astrophysical observations. The primary objective of the present work is to provide such validation by measuring pressure-broadening cross-sections for the H$_2$CO--He system at temperatures relevant to the cold ISM, and then to use the validated PES and scattering matrix to generate reliable state-to-state collisional rate coefficients for the H$_2$CO--He interaction.

Pressure broadening provides an indirect but stringent probe of intermolecular interactions. The pressure-broadening cross-section contains contributions from both elastic and inelastic collisions: elastic collisions broaden spectral lines through interference between the phase shifts accumulated by the initial and final rotational states, while inelastic collisions contribute through collision-induced transitions out of both states. The measured cross-sections correspond to thermal averages over the collision energy distribution at a given temperature \citep{wiesenfeld_ab_2010}. Because of this averaging, pressure-broadening cross-sections might be expected to exhibit only limited sensitivity to details of the PES and associated collisional propensity rules. However, recent low-temperature measurements for the HCN/HNC--He systems demonstrated that such sensitivity can remain significant even after thermal averaging \citep{hays_collisional_2022}.

The H$_2$CO--He system is of particular astrophysical interest because, although molecular hydrogen is the dominant constituent of cold molecular clouds, helium accounts for approximately 20\% of the gas by abundance. Furthermore, despite collisional excitation cross-sections with H$_2$ generally being about twice those with He, He collisions remain important and are routinely included in radiative transfer calculations alongside H$_2$ collisions (see, for example, \citet{gerin2024h2co}).

At present, collision-induced rotational (de-)excitation rate coefficients are available for H$_2$CO in collisions with He \citep{garrison1975cooling, garrison_coupled-channel_1976, green1978collisional, green1991collisional}, with both \textit{ortho}-H$_2$ and \textit{para}-H$_2$ \citep{troscompt2009rotational, Wiesenfeld_Faure_2013}, and more recently with electrons \citep{gerin2024h2co}.

Several pressure broadening measurements in the microwave region for the H$_2$CO--He system have been reported previously, although all were performed near room temperature \citep{srivastava1973microwave, rogers1973experimental, nerf1975pressure, venkatachar1975collision, feuillade1978foreign, bestmann1979investigation}. Other spectroscopic studies in the UV or IR spectral regions have also reported pressure-broadening coefficients for H$_2$CO induced by He \citep{hindmarsh_pressure_1974, barry_measurements_2003,wangIntrapulseQuantumCascade2012}. Measurements at temperatures relevant to the ISM were later carried out using a collisional cooling cell for \textit{ortho}-H$_2$CO in collisions with both He and H$_2$ \citep{mengel2000helium}. These experiments provided the first direct evidence that H$_2$ collisional cross-sections are approximately twice those for He.

Theoretically, pressure-broadening cross-sections are obtained from quantum scattering calculations based on the scattering matrix ($S$-matrix), which requires an accurate intermolecular interaction potential. The same calculations also yield state-to-state inelastic cross-sections from which astrophysically relevant collisional rate coefficients are derived. Consequently, agreement between calculated and measured pressure-broadening cross-sections provides a stringent test of both the PES and the scattering methodology, thereby increasing confidence in the resulting collisional rate coefficients.

The H$_2$CO--He rate coefficients currently adopted in astrophysical databases are those of \citet{green1991collisional}, computed using the PES of \citet{garrison1975effect} and close-coupling scattering calculations. Nearly three decades later, \citet{wheeler2003new} reported a new CCSD(T) PES that exhibited differences of up to 50\% in the well depth relative to the earlier surface, highlighting the need for renewed theoretical and experimental investigations of this system.

In the present work, we construct a new state-of-the-art PES for H$_2$CO--He and compute a corresponding set of collisional rate coefficients. The PES is validated against low-temperature pressure-broadening cross-sections measured using a recently developed experimental methodology that combines laser photolysis with uniform supersonic flows \citep{hays_collisional_2022}. We further assess the astrophysical implications of the new rate coefficients through radiative transfer calculations for astronomically observed H$_2$CO transitions, quantifying the effect of including a 20\% helium abundance on the derived excitation temperatures.

During the final stages of preparing this manuscript, we became aware of the recent work of \citet{santelices_rosas_2026}, who reported a state-of-the-art rigid-rotor PES for the H$_2$CO--He system and its deuterated isotopologues. As shown later in this paper, our interaction potential is both qualitatively and quantitatively consistent with theirs, leading to very similar collisional cross-sections and rate coefficients.

\section{Methods}
\subsection{Experimental}

\begin{figure}[h!]
   \centering
   \includegraphics[width=\hsize]{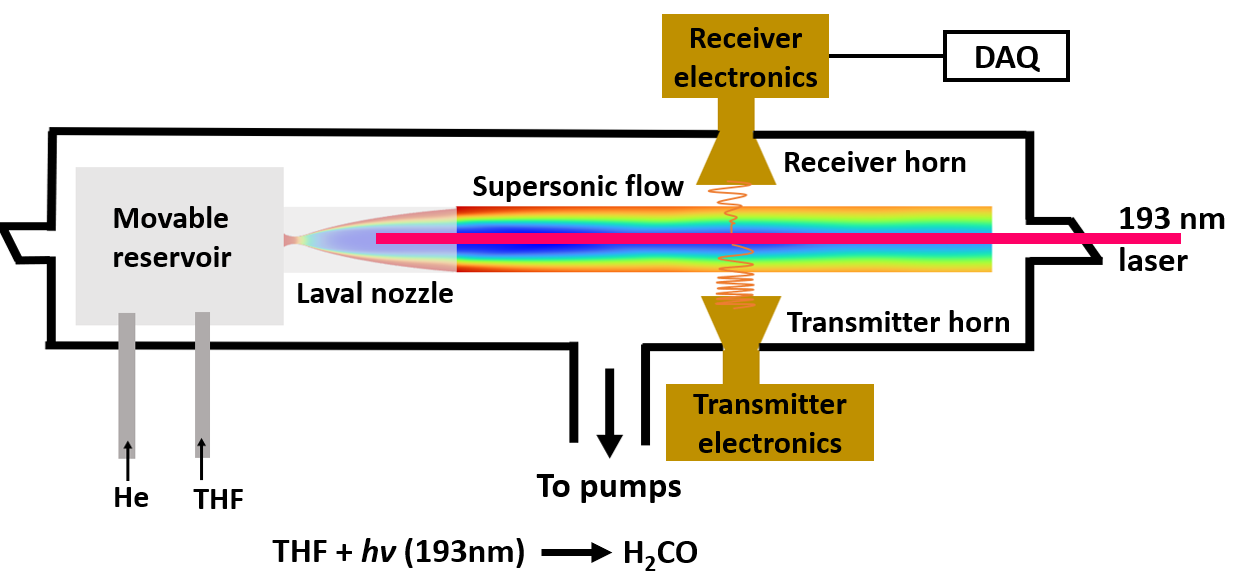}
      \caption{Schematic of the experimental setup used in deriving the low temperature pressure-broadening cross-sections.}
         \label{fig1}
   \end{figure}

A schematic of the experimental apparatus is shown in Fig.~\ref{fig1}. The experiment employs the recently implemented chirped pulse in uniform supersonic flow (CPUF) technique at Rennes \citep{hearne_novel_2020,hays_collisional_2022, guillaume_product-specific_2024}, first developed by \citet{oldham_chirped-pulse_2014} and \citet{abeysekera_chirped-pulse_2014}. This method combines chirped-pulse Fourier-transform millimeter-wave (CPFTmmW) spectroscopic detection with low-temperature, uniform supersonic flows generated using converging-diverging (Laval) nozzles. Such flows are well known as the CRESU (\textit{Cinétique de Réaction en Ecoulement Supersonique Uniforme}, a French acronym standing for reaction kinetics in uniform supersonic flow) technique \citep{Rowe1984study,Sims1994ultralow,CookeExperimentalStudiesReactivity2019,rowe_uniform_2022}.

In the present study, uniform supersonic flows using helium as the buffer gas are produced. To investigate the collisional relaxation of H$_2$CO by He, trace amounts ($<0.1\%$) of H$_2$CO could be introduced into the flow together with the He. However, this approach results in pressure-broadening cross-sections with contributions from collisions taking place in the boundary layers and in the ambient region between the horn antennas, in addition to the cold region of interest (see Fig.~\ref{fig1}). To avoid this contamination, a pulsed laser photolysis method is employed to generate H$_2$CO directly within the cold core of the flow following the previous studies of photo-generated HCN and HNC in collision with He in our laboratory \citep{hays_collisional_2022}. This is achieved through the photodissociation of tetrahydrofuran (THF) at a wavelength of 193~nm using light from a pulsed excimer laser (Coherent LPXpro 210). Oxetane and 2,3-dihydrofuran as sources for H$_2$CO under 193 nm photodissociation were also explored, but the H$_2$CO yield in the ground rovibronic state was higher for tetrahydrofuran (see Appendix \ref{AppA}).

Formaldehyde rotational transitions were detected using a CPFTmmW spectrometer covering the E-band (60-91 GHz) frequency range \citep{hays_design_2020}. In this technique, the transient emission from a molecular ensemble is recorded following the coherent excitation by a broadband microwave polarization pulse. This transient emission is detected in the time-domain, well known as the free induction decay (FID). The signal decays with time due to the loss of coherence among the molecules in the ensemble. This  decoherence is introduced mainly by the Doppler effect and molecular collisions, which correspond to the Gaussian and Lorentzian widths, respectively, of a Voigt profile in the frequency domain \citep{haekel1988determination,hays_design_2020,hearne_novel_2020}. Eq.\ref{eq1} shows the functional form of the detected signal: 
 \begin{equation}
    \label{eq1}
      S(t) = A\exp\left(-\frac{t}{T_2}\right)\exp\left(-\frac{\pi^2\Delta v_{Dopp}^2t^2}{\ln(2)}\right)\cos(2\pi vt + \theta)
   \end{equation}
where $A, \nu$, and $\theta$ represent the amplitude, frequency, and phase, respectively. The parameter $T_2$ is the collisional dephasing time related to pressure-broadened half linewidth, $\Delta$$\nu_{pres}$, by the expression $T_2 = 1/2\pi\Delta\nu_{pres}$, and $\Delta$$\nu_{Dopp}$ is the Doppler broadened half width \citep{haekel1988determination}.

Millimeter-wave excitation pulses for H$_2$CO are transmitted multiple times after each 193~nm laser pulse, and the corresponding FIDs emitted from the H$_2$CO ensemble are detected. This makes it possible to monitor the signal variation of the H$_2$CO photo-product with time. Furthermore, since photodissociation can lead to products in excited rotational and vibrational states, the time variation enables the observation of the thermalization of rotational population in the flow \citep{zaleski_time-resolved_2017}. 

Different Laval nozzles are employed to study the temperature dependence of pressure-broadening cross-sections ranging from 10 to 30~K. Each measurement is performed at a fixed pressure. Given the well-defined conditions of the flow, the pressure-broadening cross-section is obtained directly from the $T_2$ decay using Eq.\ref{eq2}
\begin{equation}
    \label{eq2}
\sigma_{PB}(T) = \frac{1}{N\overline{v}T_2} = \frac{2\pi\Delta\nu_{pres}}{N\overline{v}}
\end{equation}
where $N$ is the number density in the flow and $\overline{v}$ is the mean relative velocity at a temperature, $T$. Details of the nozzles used are tabulated in Appendix \ref{AppB}. 

For data acquisition, a 2.5~$\mu$s long pulse sequence which includes the excitation pulse and FID recording time is used. The digitizer card (Teledyne SP Devices ADQ7DC-PCIE) samples at 5~Gsa/s and has a record length of 2000000, so that 160 pulse sequences can be recorded. The data is collected for a few pulse sequences before the laser firing for background subtraction. Depending on the signal-to-noise ratio required, horizontal averaging of the sequential FIDs obtained after the laser shot is performed.

Measurements were also performed at room temperature (295~K) in the same chamber without using a nozzle, with the gas reservoir positioned far from the detection region, resulting in a subsonic flow at 295~K  at the measured chamber pressure. In this configuration, H$_2$CO was introduced directly into the chamber together with He. The H$_2$CO vapor was generated by heating paraformaldehyde, following the protocol described by \citet{ocana_fernandez_gas-phase_2020}, and was subsequently mixed with He, resulting in a gas mixture containing 3.07\% H$_2$CO. Measurements were performed at different pressures by increasing only the He flow through a separate gas line while maintaining a constant flow rate of the H$_2$CO--He gas mixture. Under these conditions, the H$_2$CO concentration in the total gas mixture remained below 0.5\%. The measurements were carried out over a pressure range of 17 to 100~$\mu$bar. The corresponding Lorentzian half-widths were obtained by fitting the time-domain spectra with the time-domain Voigt profile. The pressure-broadening coefficients, $\gamma_{\mathrm{pres}}$, were determined from the slopes of the linear dependence of the Lorentzian half-widths on pressure. Finally, the pressure-broadening cross-sections, expressed in units of $\si{\angstrom}^2$, were calculated using Eq.~\ref{eq3}.
\begin{equation}
    \label{eq3}
    \sigma_{PB}(T) = 0.596\gamma_{\mathrm{pres}}\sqrt{\mu T} 
\end{equation}
with $\mu$ being the reduced mass of the colliding species in atomic mass units, $T$ the temperature in kelvin, and $\gamma_{\mathrm{pres}}$ in MHz (mbar)$^{-1}$ units.

In these measurements, a 10~$\mu$s recording frame \citep{guillaume_product-specific_2024} was employed. The digitizer, operating at a sampling rate of 5~GSa/s, allowed the acquisition of data from 40 consecutive pulse sequences, which were subsequently averaged to improve the signal-to-noise ratio.

Formaldehyde is an asymmetric rotor species with rotational levels defined by $j_{k_a,k_c}$, where $j$ is the main rotational quantum number and $k_a,k_c$ are the projection of $j$ on the $A$ and $C$ rotational axes. It exists, due to its symmetry, in the form of two nuclear spin isomers, \textit{ortho}-H$_2$CO and \textit{para}-H$_2$CO. The nuclear spin configuration is solely defined by the $k_a$ quantum number: it must be even or zero in the case of {\it para}-H$_2$CO, while it must be odd in the case of {\it ortho}-H$_2$CO. Note also that the sum of $k_a + k_c$ must be equal to $j$ or $j+1$.
Within the spectrometer range, two transitions of formaldehyde could be reliably detected, the 1$_{0,1}$--0$_{0,0}$ transition of \textit{para}-H$_2$CO at 72837.9480~MHz and the 5$_{1,4}$--5$_{1,5}$ of \textit{ortho}-H$_2$CO at 72409.0832~MHz.

\subsection{Theoretical}

\subsubsection{Ab initio and analytical PES}

In order to carry out quantum scattering calculations, we have constructed a new high-level 3-dimensional (3D) PES, keeping the intramolecular coordinates of H$_2$CO frozen, thus neglecting its vibrations (rigid rotor approximation). The new PES has been calculated by the explicitly correlated CCSD(T)-F12b method, and the augmented correlation-consistent polarized valence triple zeta (aug-cc-pVTZ) basis set as implemented in the \texttt{MOLPRO} (version 2015.1.8) quantum chemistry software package \citep{MOLPRO_brief,Werner2012,Adler_2007}. The interaction potential uses the Jacobi coordinate system, with $R$ being the intermolecular distance between He and center of mass of H$_2$CO, as well as $\theta$ and $\phi$ polar and azimuthal angles, that define the angular orientation of He with respect to the center of mass of H$_2$CO. The internal coordinates for H$_2$CO which are $1.117$ \AA~and $1.207$ \AA~for the C--H and C--O bond lengths as well as $121.9^\circ$ and $116.2^\circ$ angles for the O--C--H and H--C--H bonds, respectively. This corresponds to the equilibrium structure that is found by full geometry optimization at the same level of coupled cluster theory. To eliminate basis set superposition errors, we applied the counterpoise procedure proposed by \citet{Boys1970}.

A broad grid with radial distances that span the range from 2 \AA~ to 26.5 \AA~(a total of 34 values of $R$, with a non-uniform step size), and an angular grid with a step size of $10^\circ$ both for $\theta$ and $\phi$ coordinates are used. In order to eliminate possible size-inconsistency effects of the CCSD(T)-F12b method, all interaction energies have been corrected by subtracting the long-range interaction energies calculated for all angular orientations at a separation of $R = 100$~bohr. Note that we use hereinafter atomic units ($a_0$) for distances ($1 a_0 =1$~bohr $\approx 0.529177$ \AA), and wavenumbers (cm$^{-1}$) for energies (1~cm$^{-1} \approx 1/219474.624$~hartree).

For the analytical representation of the PES, required for quantum scattering calculations, we employed a standard linear least-squares fit over spherical harmonic functions, as described by the following equation:
\begin{equation}
V(R, \theta, \phi) = \sum_{l}^{l_\mathrm{max}} \sum_{m}^{l} V_{lm}(R)\frac{Y_l^m(\theta, \phi) + (-1)^m  Y_l^{-m}(\theta, \phi)}{1+ \delta_{m,0}} ,
\label{eq:vrtp}
\end{equation}
where $V_{lm}(R)$ are the radial coefficients to be fitted and $Y_l^m(\theta, \phi)$  are the normalized spherical harmonics and $\delta_{m,0}$ is the Kronecker $\delta$. The indices $l$ and $m$ define the angular anisotropy (spherical harmonics degree and order, respectively). The following criteria apply: $l$ and $m$ should be positive or zero, $m$ is a multiple of 2 and less or equal to $l$ due to the $C_{2v}$ symmetry constraints of H$_2$CO. We found that including anisotropies up to $l = 15$ (leading to a total of $72$ $lm$-terms) is sufficient to ensure a good accuracy, while also maintaining efficient use and implementation of the PES in scattering codes. The accuracy of the fit is on the wavenumber level of accuracy overall and always within $3\%$ uncertainty. In particular, the mean absolute errors (MAE) and root mean square deviations (RMSD) are the largest at the short range ($R < 4$~bohr), where the interaction potential is strongly repulsive. In this region the MEA is $\sim 0.18\%$ and the RMSD is $\sim150$~cm$^{-1}$, while they quickly decrease with the distance, leading to as low as $\sim 0.015\%$ MAE and $\sim0.09$~cm$^{-1}$ RMSD at  $R = 6$~bohr, which corresponds to the intermolecular separation of the global minimum. The quality of the fit is also satisfactory in the weakly interacting long range region, for example at $R = 25$~bohr, the MEA is only $\sim 0.25\%$ and the RMSD is $\sim 5.6\times 10^{-5}$~cm$^{-1}$. To span the whole radial range needed for proper scattering calculations, a standard cubic spline interpolation of the expansion coefficients has been used for arbitrary internal distances from $R = 3.75$ to $R=25$~bohr, which was then smoothly connected to standard exponential ($V_\mathrm{sr} \approx A \exp(-\alpha R)$, with a transition domain from  $3.75$ to $4.0$~bohr). and power-law extrapolations ($V_\mathrm{lr} \approx B/R^{\beta}$ with a transition domain from $20$ to $25$~bohr) towards the short and long-range distances, respectively, using switching functions in the defined transition domains \cite[more details are given in][]{Demes_2025}.

\subsubsection{Molecular scattering calculations}
\label{section:scattering} 
For the calculation of cross-sections, we rely on accurate quantum scattering theories. Numerically exact pressure-broadening (and inelastic) cross-sections can be derived from the scattering $S$-matrices by solving the close-coupling (CC) equations on the H$_2$CO + He PES \citep{Arthurs_Dalgarno1960}. For this, we implemented the new interaction potential in the \texttt{MOLSCAT} quantum scattering code \citep{Hutson_2019} using the body-fixed reference frame, which allows the full CC treatment of the collision of a closed-shell rigid asymmetric rotor with a closed-shell atom. The coupled equations were propagated from the short to the long range using the modified log-derivative/Airy hybrid propagator of \citet{Alexander_Manolopoulos_1987}. In the asymptotic long range, the interaction potential can be neglected, and the Schr\"{o}dinger equation can be solved analytically through wavefunctions expressed by Riccati-Bessel and modified Bessel functions. This enables the derivation of the complex symmetric $S$-matrix at these boundary conditions.

In the total angular momentum representation, the complex generalized spectroscopic pressure broadening cross-sections for rotational lines can then be expressed as follows \citep{Thibault_2025, Monchick_Hunter_1986}:
\begin{equation}
\begin{split}
&\sigma_\lambda^{(q)}(j_i',j_f';j_i,j_f;E_{\rm coll}) = \frac{\pi}{k^2}(-1)^{(\lambda+j_f-j_f')} \\
&\quad\times \left(\frac{[j_i']}{[j_i]}\right)^{1/2}
\sum_{\ell,\ell',\bar\ell,\bar\ell'} \bigl([\ell][\ell'][\bar\ell][\bar\ell']\bigr)^{1/2}
\times i^{(\ell-\ell'-\bar\ell+\bar\ell')}(-1)^{\ell'-\bar\ell'} \\
&\quad \times \begin{pmatrix} \ell & \bar\ell & \lambda \\ 0 & 0 & 0 \end{pmatrix}
\begin{pmatrix} \ell' & \bar\ell' & \lambda \\ 0 & 0 & 0 \end{pmatrix}
\sum_{J,\bar J}[J][\bar J]
\times
\begin{bmatrix}
\ell & \ell' & j_f & j_f' \\
\bar\ell & j_i & \bar\ell' & j_i' \\
\lambda & \bar J & J & q
\end{bmatrix} \\
&\quad\times \Bigl[\delta_{j_i j_i'}\,\delta_{j_f j_f'}\,\delta_{\ell\ell'}\,\delta_{\bar\ell\bar\ell'} - \langle j_i \ell \,|\, S^{J}(E_{\rm coll}+E_{i}) \,|\, j_i' \ell' \rangle\,\\
&\qquad\times \langle j_f \bar\ell \,|\, S^{\bar J}(E_{\rm coll}+E_{f}) \,|\, j_f' \bar\ell' \rangle^{*} \Bigr], 
\label{eq:pbxs}
\end{split}
\end{equation}
where the notation $[X]$ corresponds to $[X] = 2X + 1$ and the different parentheses $(...)$ and $[...]$ are, respectively, Wigner 3-$j$ and 12-$j$ symbols of the second kind. The total angular momentum vector is defined by coupling the rotational angular momentum $\vec{j_i}$ of the target molecule with the total orbital angular momentum $\vec{\ell}$ of the rotating complex: $\vec{J} = \vec{j_i} + \vec{\ell}$ (the unprimed $j, \ell$ and primed $j',\ell'$ variables refer to pre- and post-collision states, respectively, and the $k_a$, $k_c$ projection quantum numbers are not involved explicitly in this formula). The spectral tensor rank $q = 0$ is for isotropic Raman lines, $q = 1$ for electric dipole lines, and $q = 2$ for anisotropic Raman or electric quadrupole lines \citep{Thibault_2025}. In the case of velocity tensor rank $\lambda = 0$, the real $\operatorname{Re}[\sigma^{(q)}_0]$ and imaginary parts $\operatorname{Im}[\sigma^{(q)}_0]$ of this cross-section describe pressure broadening and pressure shift of the spectral line, respectively. When $\lambda = 1$, the velocity vector couples with the spectral transition operator, which is the so-called Dicke-effect, but these are not calculated in the current work. $S^{J}$ are the scattering matrices for total angular momentum $J$, while $E_{i}$ and $E_{f}$ define the internal energy of the initial and final states, respectively. Note that pressure broadening calculations require $S$-matrix elements involving initial and final levels at the same kinetic energy.

Averaging the generalized spectroscopic cross-sections over a Maxwell-Boltzmann distribution of collision energies allows the derivation of the temperature-dependent quantities. The pressure-broadening cross-section functional dependence on temperature is given by:
\begin{equation}
\operatorname{Re}[\sigma^{(q)}_0(T)] = \left(\frac{1}{k_BT}\right)^2\,\int_0^{\infty} x e^{-x}\, \operatorname{Re}[\sigma_{0}^{(q)}(E_{\rm coll})]\,{\rm d}x
\label{eq:coeffs}
\end{equation}

where $k_{\rm B}$ is the Boltzmann constant, $c$ is the speed of light, $T$ is the temperature, and $E_{\rm coll} = xk_{\rm B}T$.

 In order to derive the $S$-matrix elements and the spectroscopic cross-sections, all rotational levels of H$_2$CO below $100$~cm$^{-1}$ have been targeted in our calculations, involving states up to $j=8$. We computed inelastic cross-sections for these states at collision energies between $0.1$ and $1500$~cm$^{-1}$, which is a much broader range compared to that used by \citet{santelices_rosas_2026}. The reduced mass of the H$_2$CO--He complex is 3.52941 a.m.u. We used the rotational and centrifugal distortion constants derived from the high-resolution spectroscopy measurements of \citet{Bocquet_1996} as follows (units are in cm$^{-1}$): $A = 9.4056020$, $B = 1.2954418$, $C = 1.1342006$ and $D_J = 2.5125721 \times 10^{-6}$, $D_{JK} = 4.3062377 \times 10^{-5}$, $D_K = 6.4785016 \times 10^{-4}$. A comprehensive list of rotational levels are provided in Table 1 by \citet{Wiesenfeld_Faure_2013}, which show a perfect agreement with those obtained in our scattering calculations from the spectroscopic constants. As it is discussed in this latter paper, many consecutive $j_{k_a,k_c}$ and $j_{k_a,k_c+1}$ levels at high-$j$ tend to be almost degenerate, which were correctly separated in our calculations.

Close-coupling calculations can be made more efficient by optimal truncation of some parameters, in particular the rotational basis set ($j_\mathrm{max}$) and largest total angular momentum ($J_\mathrm{tot}$). We made careful convergence tests to set these values, which does not affect the accuracy of the calculations significantly. The convergence threshold criteria were set to a maximum absolute error of $\leq 0.5\%$ for $j_\mathrm{max}$ for both elastic/inelastic cross sections, while finding a convergence on $J_\mathrm{tot}$ has been handled by the built-in convergence-test module of \texttt{MOLSCAT} with DTOL=0.3 and OTOL=0.005 parameters that allows only absolute errors not larger than $0.3~\AA^2$ for elastic and $0.005~\AA^2$ or the inelastic cross sections. Such criteria can be fulfilled only when including rotational levels up to at least $j_\mathrm{max} = 10$ at the lowest collision energies and going up to $j_\mathrm{max} = 28$ at the highest energies. The partial cross-sections from specific angular momenta are also energy-dependent, so $J_\mathrm{tot}$ criteria reached $88$ at the highest collision energies. The parameters for the numerical propagation (mixed log-derivative and Airy propagators) have been optimized with a convergence criteria of $<0.01\%$ maximum errors. The $E_\mathrm{max}$ cut-off parameter was used to exclude the non-significant contributions from high-lying rotational levels (set based on a convergence criterion $<0.1 \%$). To resolve the resonances in the low-energy cross-sections, a very fine energy grid was chosen: $0.1$~cm$^{-1}$ at energies below $50$~cm$^{-1}$ collision energy, which was gradually increased up to only a few points between the highest collision energies from 1000 to 1500 cm$^{-1}$.

The calculated $S$-matrices also allowed us to derive the inelastic cross-sections for transitions from the initial rotational state $i$ to the final rotational state $f$:
\begin{equation}
\sigma_{{i}\to {f}}(E_\mathrm{coll})  = \frac{\pi}{g_{{i}} k_{i}^2}
\sum_{J} (2J+1)
\sum_{i, f}
\left| \delta_{if} - S_{if}^{J_{\rm}}(E_\mathrm{coll}) \right|^2 ,
\label{eq:inelast}
\end{equation}
where $i$ and $f$ denote the initial and final states, $J$ is the total angular momentum, $g_{i}$ is the degeneracy of the initial level, and $k_{i} = \sqrt{2E_\mathrm{coll}\times \mu/\hbar^2}$ is the asymptotic wave vector. Note that $S_{if}^{J_{\rm}}$ are the S-matrix for inelastic transitions (${i}\to {f}$), while Eq.~\ref{eq:pbxs} involves S-matrix elements for elastic transitions in the initial ($i$) and final ($f$) states, respectively.
The corresponding state-to-state thermal rate coefficients could then be calculated:
\begin{equation}
    k_{{i} \rightarrow {f}}(T) = \left(\frac{8}{\pi\mu k_\mathrm{B}^3 T^3}\right)^\frac{1}{2} \int_{0}^{\infty} E_\mathrm{coll} \,  e^{-\frac{E_\mathrm{coll}}{k_\mathrm{B}T}} \,  \sigma_{{i} \rightarrow {f}}(E_\mathrm{coll}) \,  \mathrm{d}E_\mathrm{coll} .
    \label{eq:rates}
\end{equation}

\section{Results}

\subsection{\label{sec:experimental} Pressure-broadening measurements}
\begin{table*}[!ht]

\caption {Experimental conditions for uniform supersonic flows at the probe region with the $T_2$ decay times and the $\sigma_{PB}(T)$ for the \textit{para}-H$_2$CO 1$_{0,1}$$-$0$_{0,0}$ transition.}
\label{table:1} 
\centering
\small
\begin{tabular}{ccccccccc}
\hline\hline         
Nozzle name & Temperature & Density & THF  & No.of laser & No.of  & time range & $T_2$ & $\sigma_{PB}(T)$ \\
  & (K) & (cm$^{-3}$) & ($\%$ in the flow) & shots & FIDs (seq. avg) & ($\mu s$) & (ns) & ($\AA^{2}$) \\
\hline
\noalign{\vskip 0.5mm}
   He10K & 9.3 & 7.4$\times$10$^{16}$ &  0.05 & 5$\times$10$^{5}$ & 2 (8)  & 40-80 & 87 $\pm$ 7 & 66 $\pm$ 9.5   \\
\hline
\noalign{\vskip 0.5mm}
   He15K &14.8 & 5.0$\times$10$^{16}$ &  0.07 & 2.5$\times$10$^{5}$ & 7 (4) & 10-80 & 121 $\pm$ 3 & 55 $\pm$ 6 \\
\hline
\noalign{\vskip 0.5mm}
  He20K & 18.7 & 3.8$\times$10$^{16}$ &  0.09 & 7$\times$10$^{5}$ & 3 (8) & 5-80 & 129 $\pm$ 8 & 61 $\pm$ 8 \\
\hline
\noalign{\vskip 0.5mm}
  He35K &30.2 & 3.0$\times$10$^{16}$ &  0.06 & 1$\times$10$^{6}$ & 2 (8) & 40-85 & 169 $\pm$ 30 & 48 $\pm$ 10 \\
\hline
\end{tabular}
\tablefoot{Errors correspond to 95$\%$ confidence interval. A 10$\%$ uncertainty in the temperature and density of the flow are considered. Laser fluences used are between $2-6.5$ mJ/cm$^2$.}
\end{table*}

\begin{table}[!ht]

\caption {Room temperature pressure-broadening coefficients and cross-sections. Errors correspond to 95$\%$ confidence interval.}

\label{table:2} 
\centering
\small
\begin{tabular}{cccccc}
\hline\hline             
Molecule & Transition & $\gamma_{pres}$  & $\sigma_{PB}(T)$ \\
&  & (MHz (mbar)$^{-1}$) & ($\AA^{2}$) \\
\hline 
  \textit{para}-H$_2$CO  & 1$_{0,1}$--0$_{0,0}$ &  1.76 $\pm$ 0.02    & 33.8 $\pm$ 0.4  &    \\
\hline
  \textit{ortho}-H$_2$CO & 5$_{1,4}$--5$_{1,5}$ & 1.77 $\pm$ 0.04   & 34 $\pm$ 1 &  \\
\hline
\end{tabular}
\end{table}

\begin{figure}[!ht]
   \centering
   \includegraphics[width=\hsize]{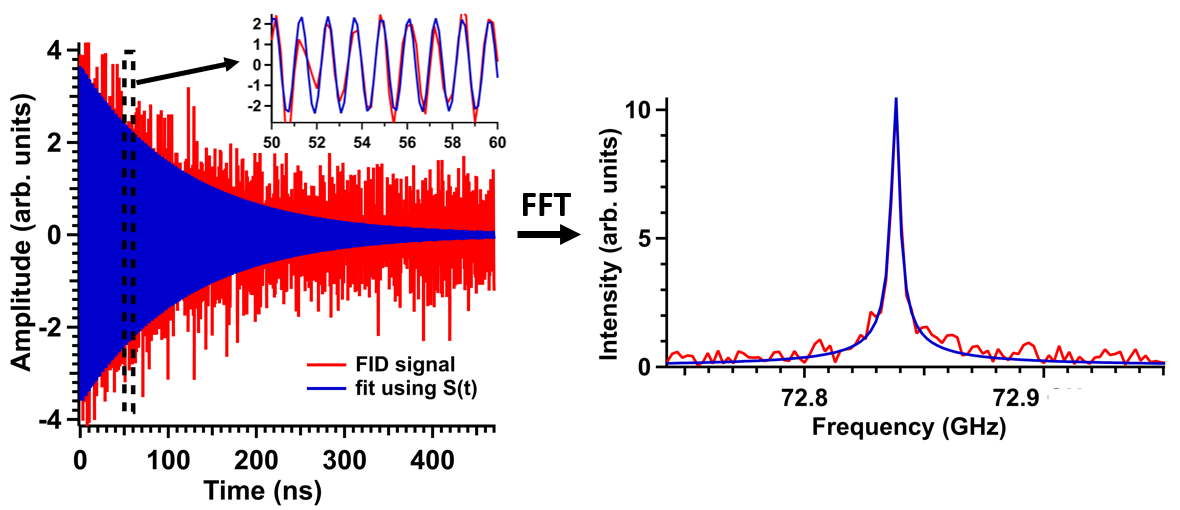}
      \caption{A FID signal of the \textit{para}-H$_2$CO transition 1$_{01}$--0$_{00}$ at 72837.9480 MHz is shown in the left panel (red), recorded in a He flow at 14.9 K. The fitted curve (blue) is overlaid, and the inset presents a magnified view of the highlighted region. The corresponding FFT magnitude is displayed in the right panel for comparison.}
         \label{FID_zoom}
   \end{figure}

Fig. \ref{FID_zoom} presents an example of a FID signal for \textit{para}-H$_2$CO generated through the photodissociation of THF in the isentropic core of a 14.9~K He flow. The decay is fitted in the time domain using Eq.~\ref{eq1}. The corresponding frequency domain signal, obtained via the Fast Fourier Transform (FFT) method, is also shown in Fig. \ref{FID_zoom}. In all cases, the Doppler broadening width is fixed to the value calculated from Eq.~\ref{eq:Doppler}. The fit well represents the obtained FID signal.

\begin{equation}
   \label{eq:Doppler}
    \Delta v_D = v\sqrt{\frac{2k_BT\ln(2)}{mc^2}}
\end{equation}

\begin{figure}[!ht]
   \centering
   \includegraphics[width=\hsize]{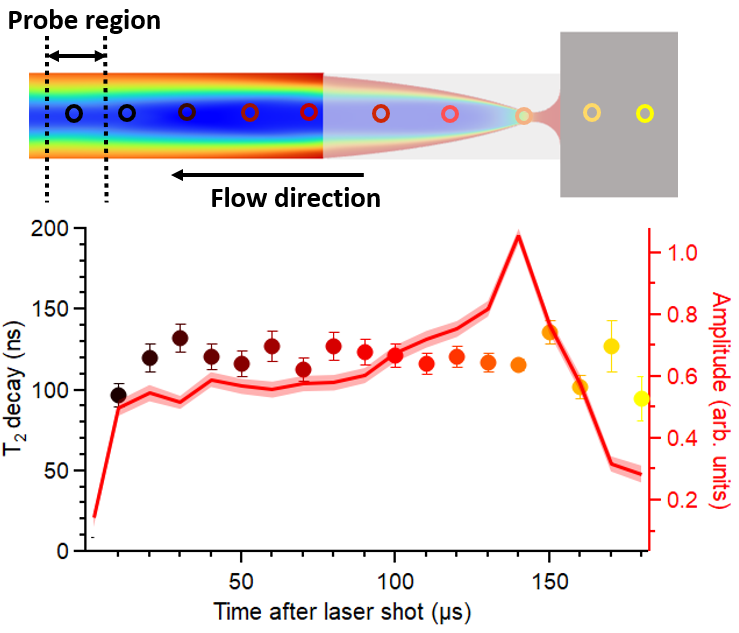}
      \caption{ The amplitude and $T_2$ decay of \textit{para}-H$_2$CO generated in a 14.9 K helium flow, plotted against the post-laser pulse delay. As depicted in the schematic above the plot, the signal captured at any specific delay represents molecules sampled from distinct spatial regions within the flow. 80 frames of data after the laser shot are recorded and 4 consecutive frames are averaged together. A total of 2.5$\times$10$^6$ laser shot-pulse sequence trains are averaged together to yield the displayed signal.}
         \label{T2_decay_time}
   \end{figure}

Multiple FIDs are recorded following each laser shot, from which the corresponding $T_2$ decay constants are extracted. These values are plotted together with the H$_2$CO signal amplitude in Fig. \ref{T2_decay_time} where four consecutive FIDs are ``horizontally'' averaged. The signal observed at a given delay after the laser shot corresponds to the arrival of H$_2$CO molecules, produced in a specific region of the flow, at the probe region. The peak in the H$_2$CO signal amplitude corresponds to molecules produced near the throat of the nozzle, where the density is relatively higher than in other regions. Using the well-defined flow velocity and the known position of the nozzle relative to the probe region, the time window corresponding to the uniform flow region can be determined. All $T_2$ decays within this interval are then averaged. H$_2$CO molecules produced inside the nozzle enter the boundary layers of the flow by the time the gas reaches the probe region and are therefore not considered in the analysis. 

The $T_2$ decay time is expected to increase for molecules produced inside the nozzle or at the throat. This is because H$_{2}$CO molecules formed inside the nozzle are likely to be present in the boundary layers, where the pressure-broadening cross-section decreases with increasing temperature. Such an increase in the $T_2$ decay time was observed in previous studies of HCN/HNC collisions with He \citep{hays_collisional_2022}. However, in the present case, the $T_2$ decay time appears to remain approximately constant within the experimental error limits as molecules produced inside the nozzle arrive at the detection zone. The intensity of the fundamental rotational transition of H$_{2}$CO  displays a stronger temperature dependence than that of HCN (or HNC), resulting in a stronger positive discrimination for the cold isentropic core region of the flow.

Uniform supersonic flows at different temperatures are obtained with different Laval nozzles, whose Pitot (impact pressure) measurements are reported in Appendix \ref{AppB} and were employed in the previous studies \citep{hays_collisional_2022, guillaume_product-specific_2024}. Table \ref{table:1} reports the flow conditions, data acquisition and analysis details and the corresponding $T_2$ decay times, alongside the derived pressure-broadening cross-sections. In Table \ref{table:1}, the temperatures and densities quoted correspond to the probe region, rather than the average value for the entire uniform flow. All the analysis was performed considering a Doppler half-width half-maximum fixed at a value determined by Eq.~\ref{eq:Doppler}.

\begin{figure}[!ht]
   
    \centering
    
    \includegraphics[width = \linewidth]{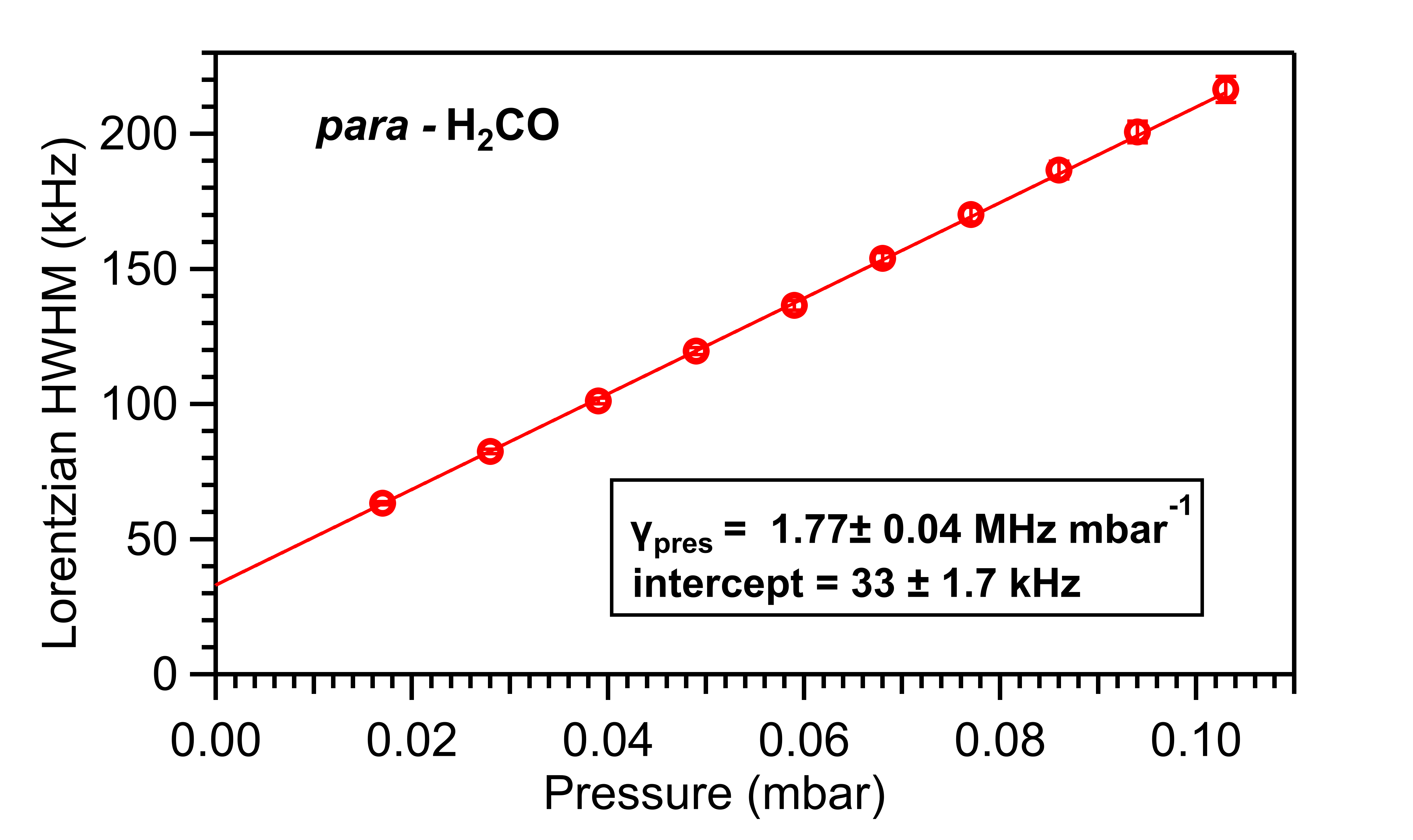}
    \includegraphics[width = \linewidth]{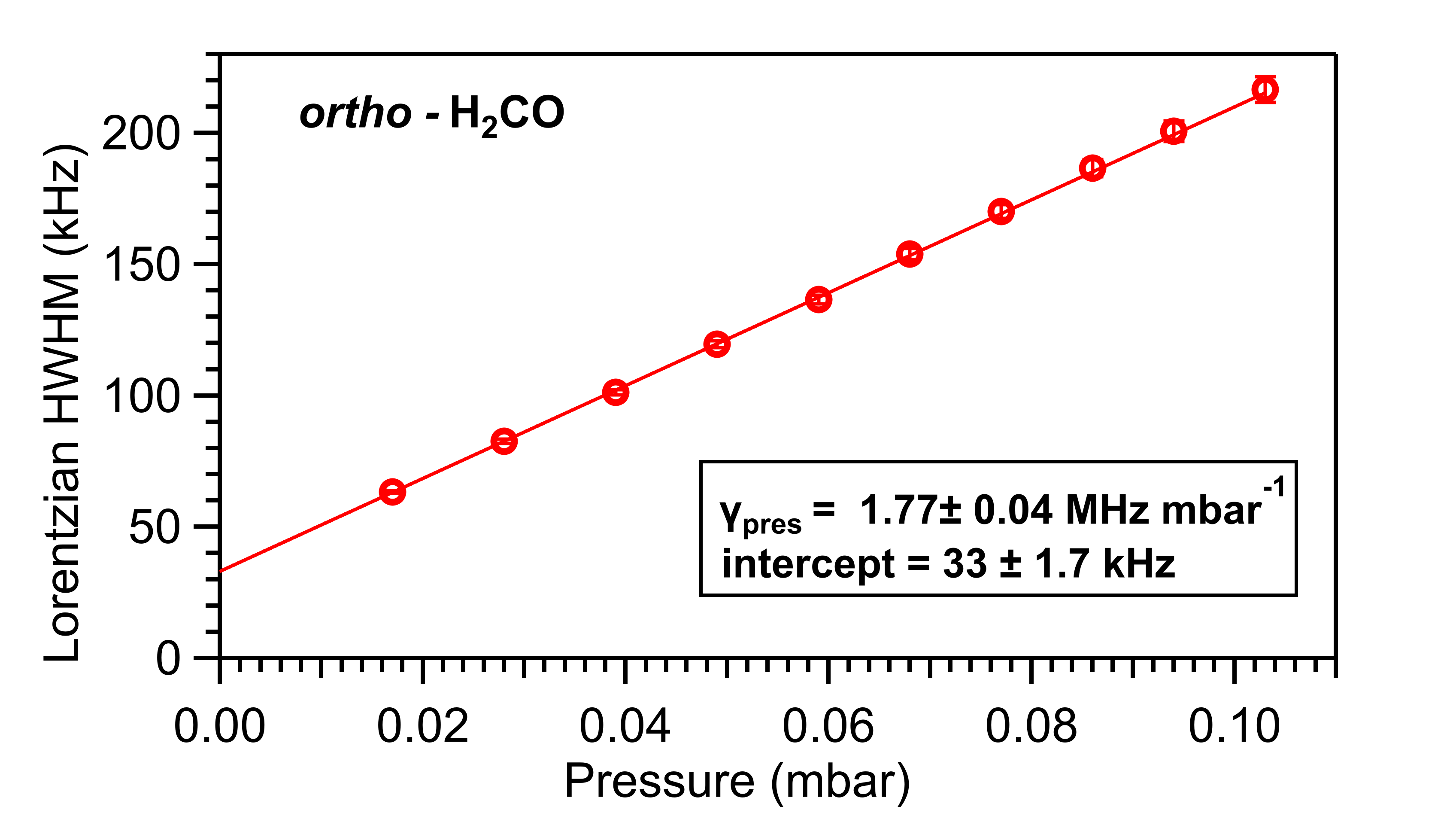}
    \caption{Plot of $\Delta\nu_{pres}$ obtained by fitting the FID to the time domain Voigt profile for the \textit{para}-H$_2$CO transition 1$_{0,1}$$-$0$_{0,0}$ (top) and the \textit{ortho}-H$_2$CO transition 5$_{1,4}$$-$5$_{1,5}$ (bottom) at different pressures and at 295 K. The slope of this plot gives $\gamma_{pres}$ reported with the 95$\%$ confidence interval.}
     \label{RT1}
\end{figure}

Low temperature pressure-broadening measurements were not achieved for the \textit{ortho}-H$_2$CO transition 5$_{1,4}$$-$5$_{1,5}$ due to insufficient signal-to-noise level at low temperatures. They are, however, possible at  295~K. Therefore, room temperature measurements are reported for the both transitions. Lorenztian HWHM (half width half maximum) values determined at different pressures are presented in Fig. \ref{RT1} for both \textit{para}-H$_2$CO and \textit{ortho}-H$_2$CO transitions. The pressure-broadening coefficients and cross-sections are reported in Table \ref{table:2}.

\subsection{PES and Scattering calculations}
\begin{figure*}[!htb]
    \centering
    \includegraphics[width=0.45\linewidth]{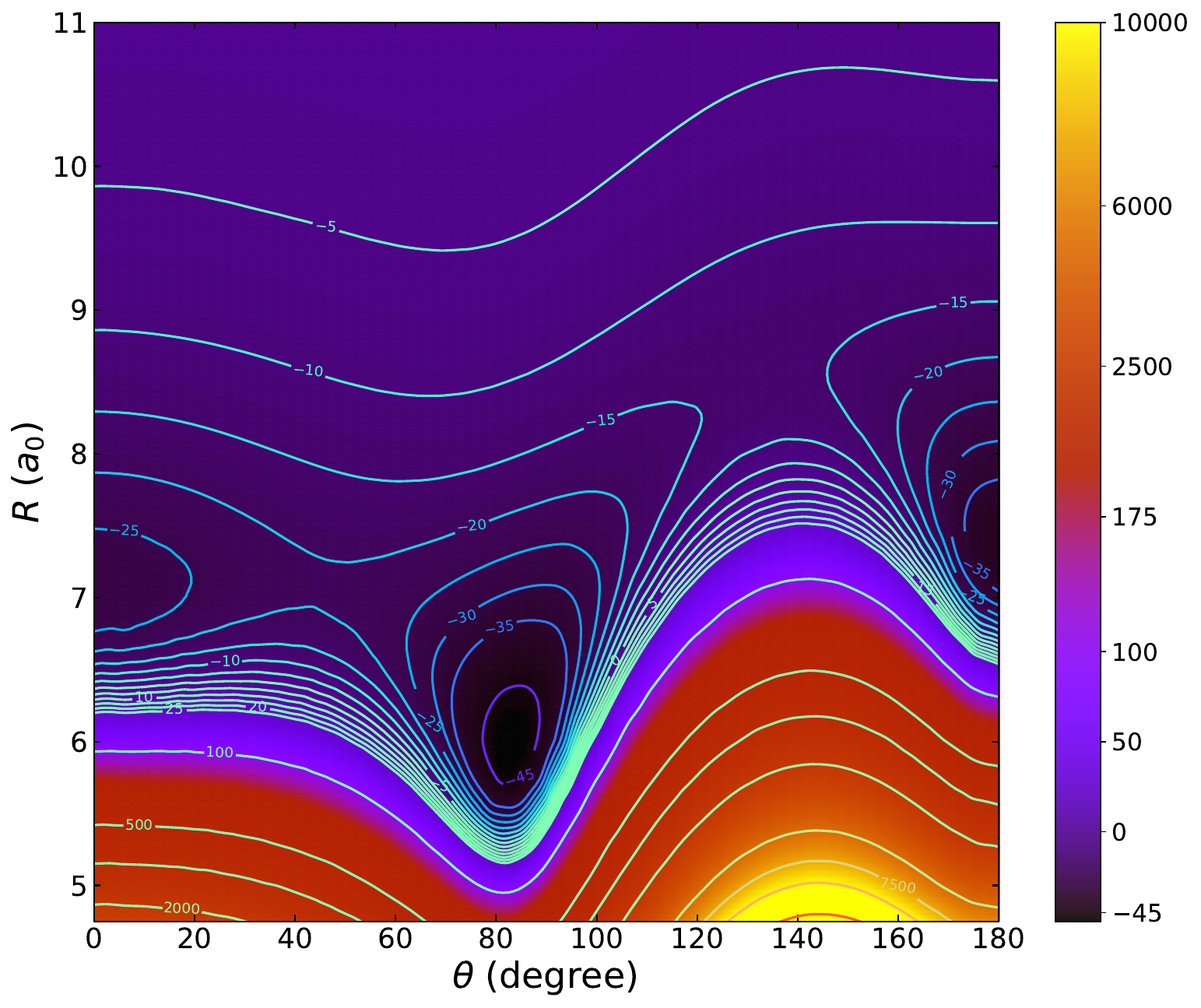}
    \includegraphics[width=0.45\linewidth]{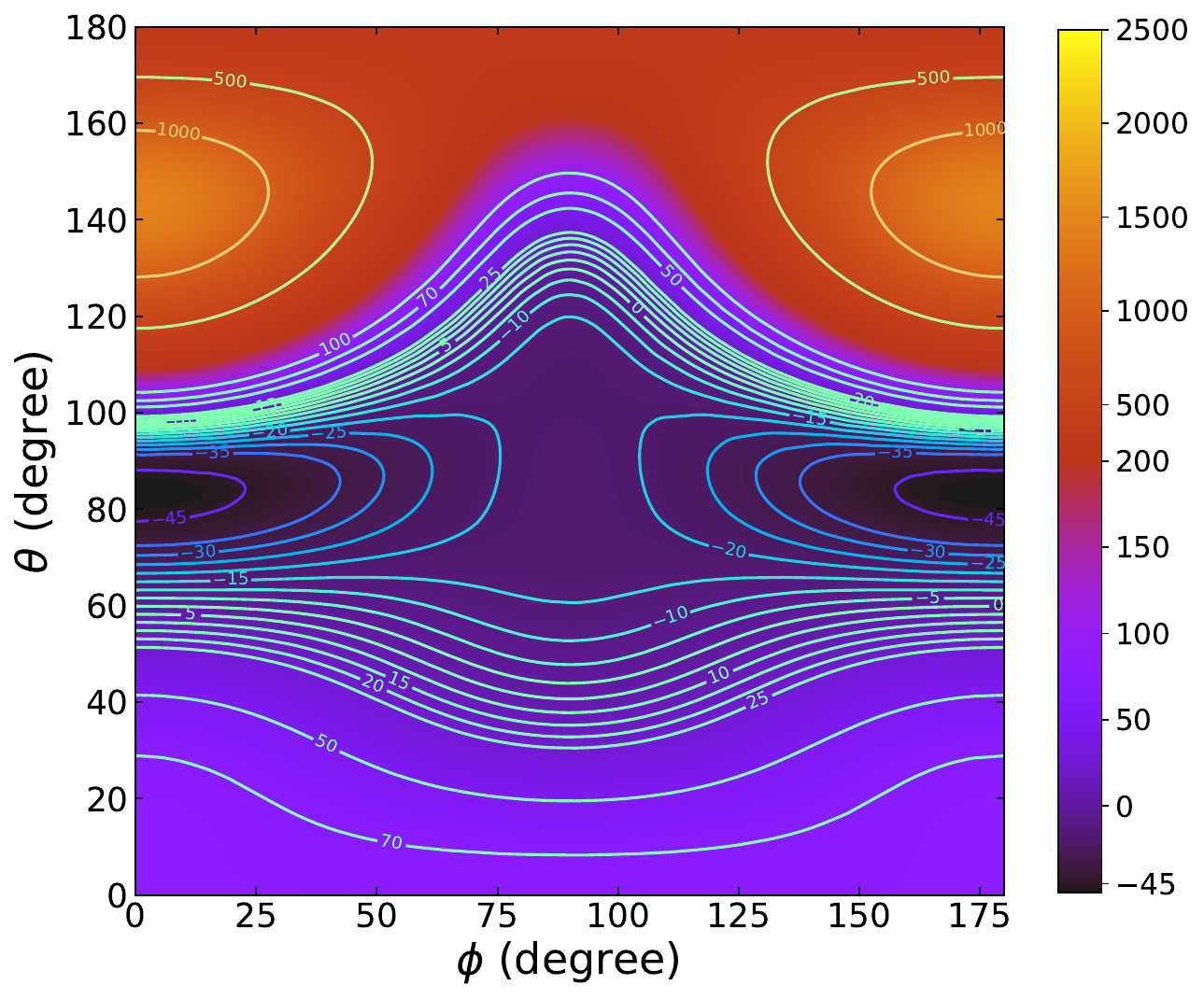}
    \caption{Contour plots for the H$_2$CO + He PES at fixed azimuthal angle $\phi=0^{\circ}$ (left) and intermolecular distance $R=6.0$~Bohr (right). The contour values refers to interaction energies in cm$^{-1}$.}
     \label{fig:PEScontours}
\end{figure*}
For a visual illustration of the PES features, we show some contour plots in Fig.~\ref{fig:PEScontours} constructed  from the analytical fit. The plot parameters are chosen to show the vicinity of the global minimum and to allow a comparison with previous studies by \citet{santelices_rosas_2026} and \citet{wheeler2003new}. The left panel of Fig.~\ref{fig:PEScontours} shows the radial dependence {\it versus} the spherical angle $\theta$ at constant $\phi=0^{\circ}$. The global well is found to be around $R \simeq 6.00$~$a_0$, $\theta \simeq 83.2^{\circ}$ and $\phi = 0^{\circ}$, which is in an excellent agreement with the recent results of \citet{santelices_rosas_2026}, that is $R \simeq 5.95$~$a_0$, $\theta \simeq 83^{\circ}$ and $\phi = 0^{\circ}$, while the well depth of our PES is somewhat shallower at $-50$~cm$^{-1}$ compared to $-53$~cm$^{-1}$ for \citeauthor{santelices_rosas_2026}. Overall, the radial dependence is rather strong and characterized with large energy gradients. A relatively deep secondary minimum is also found at $R \simeq 7.4$~$a_0$ and $\theta = 180^{\circ}$ with similar shape and width as the global one.
The right panel of Fig.~\ref{fig:PEScontours} shows the $\theta$ contour against $\phi$ angles at fixed distance $R = 6.0$~$a_0$. As expected, the PES is mostly attractive at this $R$-distance. Due to the $C_{2v}$ symmetry of H$_2$CO, there is a left-right symmetry with respect to azimuthal angle $\phi=90^{\circ}$. Respectively, {\it ab initio} calculations are only carried out in a limited interval $[0^{\circ};90^{\circ}]$, whereas the analytical fit should cover the whole sphere. This plot is therefore useful to examine, to determine whether the fitted PES correctly recovers the symmetry with respect to $\phi$. As one can see, the analytical potential describes perfectly this symmetric behavior. Note that the turning point around $\theta \simeq 83^{\circ}$ and $\phi=90^{\circ}$ has a saddle point character with a height of $\sim 30$~cm$^{-1}$ which might be mostly important at low collision energies Another noticeable aspect is that the angular anisotropy with respect to $\phi$ is rather weak compared to that in $\theta$, which means that the number of $m$-terms for a given $l$ in the analytical PES (see Eq.~\ref{eq:vrtp}) can be somewhat reduced without losing significant accuracy. Overall, the isocontour lines in Fig.~\ref{fig:PEScontours} exhibit a smooth behavior, which reflects the good quality of the fit. It is also important to mention that only minor differences are seen between our PES and the one calculated by \citet{santelices_rosas_2026}. In particular, we used the b-version of the CCSD(T)-F12 method (rather than the a-version they used), and we applied a somewhat broader radial and denser angular grid in our {\it ab initio} calculations. The slight differences are probably related to these features.

\begin{figure}[!ht]
   \centering
   \includegraphics[width=\columnwidth]{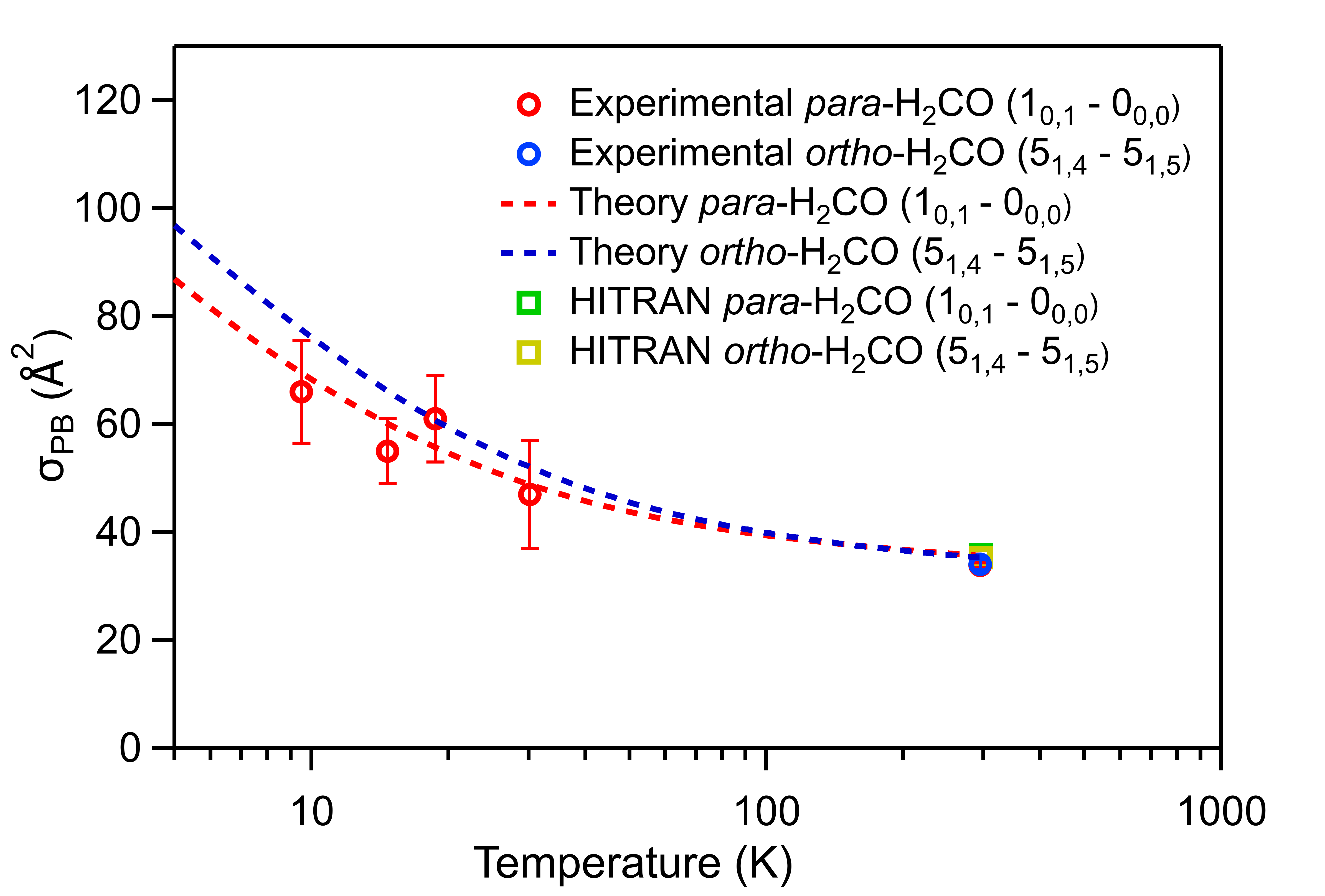}
      \caption{Pressure-broadening cross-sections measured at different temperatures are compared with the theoretical values showing an agreement within the 95$\%$ confidence interval limits of the experiments. Room temperature (296~K) values from the HITRAN database \citep{gordon_hitran2024_2026} are also shown.}
         \label{PB_xsec}
   \end{figure}

The close-coupling scattering calculations performed on this new PES yielded the energy dependent pressure-broadening cross-sections. These are presented in Appendix~\ref{AppC} and Fig.~\ref{fig:XSplot} for the two transitions under consideration. Thermal averaging then enabled the calculation of temperature-dependent pressure-broadening cross-sections which are compared with the experimental values measured at different temperatures in Fig. \ref{PB_xsec}. Note that, \textit{q}=1 case is considered in the calculation of Eq. \ref{eq:pbxs} corresponding to current experiments.

\begin{table*}[!htp]
\caption{Rotational de-excitation rate coefficients at various kinetic temperatures for the transitions from the low-lying $3_{03}$ level of {\it para}-H$_2$CO as compared to the corresponding collisional data calculated by \citet{santelices_rosas_2026} and \citet{green1991collisional}}
\label{table:rates} 
\centering
\small
\begin{tabular}{cccccc}
\hline
\hline
Transition & $T = 10$~K & $T = 20$~K  & $T = 30$~K & $T = 40$~K & Reference \\
\hline
\noalign{\vskip 0.5mm}
$3_{0,3} \rightarrow  0_{0,0}$ & $8.99 \times 10^{-12}$ & $7.25 \times 10^{-12}$ & $6.33 \times 10^{-12}$ & $5.86 \times 10^{-12}$ & This work \\
              & $8.47 \times 10^{-12}$ & $6.99 \times 10^{-12}$ & $6.17 \times 10^{-12}$ & $5.76 \times 10^{-12}$ & \citet{santelices_rosas_2026}\\
              & $3.02 \times 10^{-12}$ & $2.8 \times 10^{-12}$  & $2.77 \times 10^{-12}$ & $2.84 \times 10^{-12}$ & \citet{green1991collisional}\\
\hline
\noalign{\vskip 0.5mm}
$3_{0,3} \rightarrow  1_{0,1}$ & $4.13 \times 10^{-11}$ & $3.76 \times 10^{-11}$ & $3.54 \times 10^{-11}$ & $3.41 \times 10^{-11}$ & This work \\
               & $4.09 \times 10^{-11}$ & $3.78 \times 10^{-11}$ & $3.59 \times 10^{-11}$ & $3.48 \times 10^{-11}$ & \citet{santelices_rosas_2026}\\   
               &  $1.23 \times 10^{-11}$  & $1.36\times 10^{-11}$  & $1.52 \times 10^{-11}$  & $1.68 \times 10^{-11}$ & \citet{green1991collisional}\\
\hline
\noalign{\vskip 0.5mm}
$3_{0,3} \rightarrow  2_{0,2}$ & $5.97 \times 10^{-11}$ & $5.14 \times 10^{-11}$ & $4.64 \times 10^{-11}$ & $4.33 \times 10^{-11}$ & This work \\
              & $6.29 \times 10^{-11}$ & $5.23 \times 10^{-11}$ & $4.73 \times 10^{-11}$ & $4.42 \times 10^{-11}$ & \citet{santelices_rosas_2026}\\
              & $5.28 \times 10^{-11}$ & $5.08 \times 10^{-11}$ & $4.87 \times 10^{-11}$ & $4.68 \times 10^{-11}$ & \citet{green1991collisional}\\
\hline
\end{tabular}
\end{table*}
\begin{table}[!ht]
\centering
\caption{Percentage difference (\%~diff.) in excitation temperatures calculated with and without the inclusion of 20\% He relative abundance for transitions in warm regions.}
\label{table:Tex}
\small
\begin{tabular}{c c c c c}
\hline
\hline
Transition &
$T_{kin}$~(K)  &
\multicolumn{2}{c}{$T_{ex}$~(K)} &
\% diff. \\
&  &  Model 1 & Model 2 & \\
&  & 100\% H$_2$ & 80\% H$_2$ + 20\% He & \\
\hline

$2_{1,2}\rightarrow1_{1,1}$ & 50  & 115.59   & 114.265 & 1.1 \\
\hline
$2_{1,2}\rightarrow1_{1,1}$ & 100  & --125.017 & $-$109.934 & 12.1 \\
\hline
$3_{0,3}\rightarrow2_{0,2}$ & 50  &  26.498   & 24.729 & 6.7 \\
\hline
$3_{0,3}\rightarrow2_{0,2}$ & 100 &  51.136   & 46.270 & 9.5 \\
\hline
$3_{2,2}\rightarrow2_{2,1}$ & 50  &  28.873   & 27.047 & 6.3 \\
\hline
$3_{2,2}\rightarrow2_{2,1}$ & 100 &  66.485   & 59.659 & 10.3 \\
\hline
$4_{1,4}\rightarrow3_{1,3}$ & 50  &  6.203    & 5.843 & 5.8 \\
\hline
$4_{1,4}\rightarrow3_{1,3}$ & 100 &  30.440   & 27.654 & 9.2 \\
\hline

\end{tabular}
\tablefoot{The density considered is $1\times10^6$ cm$^{-3}$.}
\end{table}
\section{Discussion}
Excellent agreement between the theoretical and experimental pressure-broadening cross-sections can be observed in Fig. \ref{PB_xsec}. As these quantities are expressed on independent absolute scales, their comparison provides an solid basis for the proper experimental validation of both the quality of the PES and the accuracy of the scattering calculations. Therefore, one of the main achievements of the current work is the experimental cross-validation of the theoretical methodology that is commonly used to compute collisional rate coefficients for astrophysical applications and also to assess its accuracy in different temperature regimes. In such calculations, the most uncertain, hardly quantifiable errors typically arise from the PES: both its {\it ab initio} calculation and its analytical fit (especially in relation to  short- and long-range extrapolations). The excellent agreement between experiments and theory in principle validate both the methodologies.

Previous low-temperature pressure-broadening measurements by \citet{mengel2000helium}, performed using a collisional cooling cell, were reported for the $2_{1,2}$--$1_{1,1}$, $2_{1,1}$--$1_{1,0}$, and $3_{1,3}$--$2_{1,2}$ transitions, which differ from those investigated in the present study. Nevertheless, between 10~K and 30~K, the pressure-broadening cross-sections for the $1_{0,1}$--$0_{0,0}$ transition are consistent, within the experimental uncertainties, with those reported for the other transitions, suggesting that pressure-broadening cross-sections exhibit little dependence on the rotational transition for levels with $j \leq 3$. At 299 K, a pressure-broadening coefficient of 1.28 MHz/mbar was obtained for the 5$_{1,4}$--5$_{1,5}$ transition by \cite{rogers1973experimental}. No other studies exist, to our knowledge, for the transitions investigated in this work. At 296~K, pressure-broadening coefficients are reported in the HITRAN database as 1.86 MHz/mbar and 1.83 MHz/mbar respectively for the $1_{0,1}$--$0_{0,0}$ {\it para}-H$_2$CO and $5_{1,4}$--$5_{1,5}$ {\it ortho}-H$_2$CO transitions \citep{gordon_hitran2024_2026}. These values are derived by \citet{tan_h2_2022} from a semi-empirical model based on the Pad\'e approximation using measurements from \citet{nerf1975pressure,barry_measurements_2003, wangIntrapulseQuantumCascade2012}.

\section{The excitation of H$_2$CO in molecular clouds}
\label{sec:radiative} 

The validation of the PES and scattering calculations by comparison of the theoretically derived $\sigma_{\textrm{PB}}$ with the experimental values on an absolute scale enable the use of these same calculations to derive reliable state-to-state inelastic cross-sections and thermal rate coefficients for H$_2$CO--He collisions. We have compared our results with those calculated very recently by \citet{santelices_rosas_2026}, as well as those of \citet{green1991collisional}, used up until now in radiative transfer calculations \citep{gerin2024h2co}. As one can see in Table~\ref{table:rates}, there is a good overall agreement between the two most recent sets of collisional data. Relatively small differences are very likely due to the slightly different PES and the different truncation of the rotational basis used in the scattering calculations. However, significant differences exist between these recent calculations and those of \citet{green1991collisional}, mainly explained by the use of much more accurate methods for determining the H$_2$CO--He PES. In contrast to the findings of \citet{Wiesenfeld_Faure_2013} for H$_2$CO--H$_2$ collisions, the rate coefficients resulting from helium collisions exhibit only a mild dependence on temperature.

Our rate coefficients have been used to perform non-LTE radiative transfer simulations in order to determine their impact in astrophysical applications. These were performed with the \texttt{RADEX} code \citep{van2007computer} using the escape probability formalism approximation. We computed the excitation temperature ($T_{ex}$) for all transitions covered by the present set of rate coefficients. Here, however, we focus on the transitions that have been observed astronomically. Specifically, we present $T_{ex}$ for the H$_2$CO emission lines $2_{1,2} \to 1_{1,1}$, $3_{0,3} \to 2_{0,2}$, $3_{2,2} \to 2_{2,1}$, and $4_{1,4} \to 3_{1,3}$, emission lines of H$_2$CO that are frequently observed in interstellar molecular clouds \citep{troscompt_constraining_2009,Spezzano:25,Maret:04,Evans:23}.

In our calculations, the cosmic microwave background used as background radiation field was set to 2.73 K. The kinetic temperature \textit{(T$_{kin}$)} was fixed at 10 K to simulate the physical conditions of cold molecular clouds, and at 50 K and 100 K to reach the temperature of low-mass protostars and hot corinos. 

Our goal was to check the impact of considering or not He as a projectile in the radiative transfer calculations. Indeed, in molecular clouds, He accounts for $\sim$20\% of the collisions while H$_2$ accounts for $\sim$80\%. Usually, when H$_2$-rate coefficients are available, He collisions are neglected \citep{roueff_molecular_2013}. Such an approach is not expected to strongly influence the excitation conditions when He and H$_2$ collisional data agree, but it can have non-negligible effects when the two sets of collisional data differ significantly.

Two sets of calculations were performed: the first set of calculations considered H$_2$ as the only collider (model 1) while the second set of calculations considered both He and H$_2$ as colliders (with H$_2$:He ratio of 4, model 2).
Furthermore, the OPR of H$_2$ was fixed in both sets of calculations at a value corresponding to a thermal distribution of $ortho$- and $para$-H$_2$. We selected two values of the total density of molecular hydrogen $n$(H$_2$), 3$\times$10$^4$ and 10$^6$ cm$^{-3}$, corresponding to the typical density in cold molecular clouds and protostars, respectively. The column density of H$_2$CO was set at 10$^{13}$ cm$^{-2}$. 

For the calculations corresponding to cold molecular clouds, we found (not shown here) that including or not He collisions for modelling cold molecular clouds has a minor impact on the excitation conditions of H$_2$CO. In such molecular clouds, H$_2$ is mostly in its $para$ form, and as the He and $para$-H$_2$ rate coefficients agree reasonably well, models 1 and 2 were expected to agree. Hence, it seems that considering or not the impact of He collisions in radiative transfer models will not significantly change the observational analysis of cold molecular clouds.

On the other hand, including He collisions for modelling protostars and hot corinos leads to visible changes in the excitation of H$_2$CO. Table \ref{table:Tex} presents the results of radiative transfer calculations for physical conditions corresponding to warm molecular clouds.
The deviations between the two sets of calculations are larger than 10\% for some lines. Such deviations can be simply explained by the fact that He and $ortho$-H$_2$ rate coefficients differ significantly ($ortho$-H$_2$ becomes more and more abundant with increasing temperatures). It seems then important to include the He colliders when modelling warm ($T \ge 50$~K) molecular clouds.

\section{Conclusions}
Modelling astronomical observations from telescopes in non-LTE environments within the interstellar medium requires collisional rate coefficients involving the most abundant species, as the derived physical properties depend heavily on such molecular properties. Experimental methods provide an essential means to assess the uncertainty in theoretical calculations and to validate the underlying methodology. By integrating uniform supersonic flows with CP-FTmmW spectroscopy and producing molecules \textit{in situ} in the isentropic core via pulsed laser photolysis, pressure-broadening cross-sections were successfully obtained for H$_2$CO collisions with He at temperatures of relevance for the ISM (down to 10 K). A new PES was computed from high-level \textit{ab initio} theories and fitted for the H$_2$CO-He system, and then used in close-coupling scattering calculations to provide pressure-broadening cross-sections as well as collisional rate coefficients. Excellent agreement on an absolute basis was obtained between the measured and calculated pressure-broadening cross-sections, providing validation of the PES calculations and scattering methodology. By extension, this work also serves as a validation of the recently published work of \citet{santelices_rosas_2026} as their collisional rate coefficients are in very good agreement with our values. Furthermore, as this study is only the second application of this new experimental methodology, agreement between theory and experiments also increases confidence in the experimental method.  Considering helium abundances of up to 20\% relative to H$_2$, neglecting He can lead to variations in the excitation temperature of up to 9--12\%. The corresponding rate coefficients derived using the present state-of-the-art methodology will be made available in the \href{https://emaa.osug.fr/}{EMAA} \citep{Faure_2025}, \href{https://basecol.vamdc.eu/}{BASECOL} \citep{Dubernet_2024} and \href{https://home.strw.leidenuniv.nl/~moldata/}{LAMDA} \citep{vanderTak_2020} databases.

\begin{acknowledgements}
      We acknowledge CEA/GENCI for awarding us access to the TGCC/IRENE HPC facility within the A0150413001 project, and the Digital Government Development and Project Management Ltd. for awarding us access to the Komondor HPC based in Hungary. SD acknowledges financial support from the Horizon Europe Marie Slodowska-Curie Actions programme under the Grant Agreement No. 101244231 (VIBREAC). This project has received financial support from the French National Research Agency (ANR) under grant agreement ANR-24-CE30-3044 (project BRIGHTER) and from the CNRS MITI interdisciplinary programmes through its exploratory research programme. We thank Brian Hays and Alberto Macario for helpful advice and discussions.
\end{acknowledgements}

\bibliographystyle{aa}
\bibliography{references}  

\begin{appendix}

\onecolumn
\section{\label{AppA} Formaldehyde from different sources}
To avoid the contribution from the boundary layers, H$_2$CO is produced in the isentropic core region of the flow from 193 nm pulsed excimer laser photodissociation using three different molecules. A set of experiments were performed in a slit-jet supersonic expansion at a temperature of approximately 7~K in the probe region determined through vinyl cyanide transitions. The H$_2$CO is detected via the fundamental \textit{para-} transition at 72.837 GHz. The yield of the H$_2$CO from three different molecules is given in Table \ref{tb:A}. Among them tetrahydrofuran gives higher yield of H$_2$CO at the detected transition and is therefore used for the pressure-broadening experiments.
\begin{table*}[!ht]

\caption {The yield of H$_2$CO in the ground rovibronic state from different sources explored in this work.
\newline 
\textit{*signal intensity normalized to the amount of the precursor flow}.}
\centering
\large
\begin{tabular}{cccccc}
\hline\hline 
\noalign{\vskip 0.5mm}
Molecule & Absorption cross-section at 193 nm (cm$^2$) & H$_2$CO yield (arb. units)* \\
\hline 
\noalign{\vskip 0.5mm}
   Tetrahydrofuran & 2.3$\times$10$^{-18}$ &  2031    \\
\hline
\noalign{\vskip 0.5mm}
  2,3-dihydrofuran & 1.5$\times$10$^{-17}$ &  364   \\
\hline
\noalign{\vskip 0.5mm}
  Oxetane & 2.7$\times$10$^{-18}$ &  245    \\
\hline
\label{tb:A}
\end{tabular}
\tablefoot{Photoabsorption cross-sections were taken from MPI Mainz database. }
\end{table*}

\section{\label{AppB} Pitot characterisation of the Laval nozzles used}
\begin{figure}[!ht]
\centering
    \includegraphics[width=\linewidth]{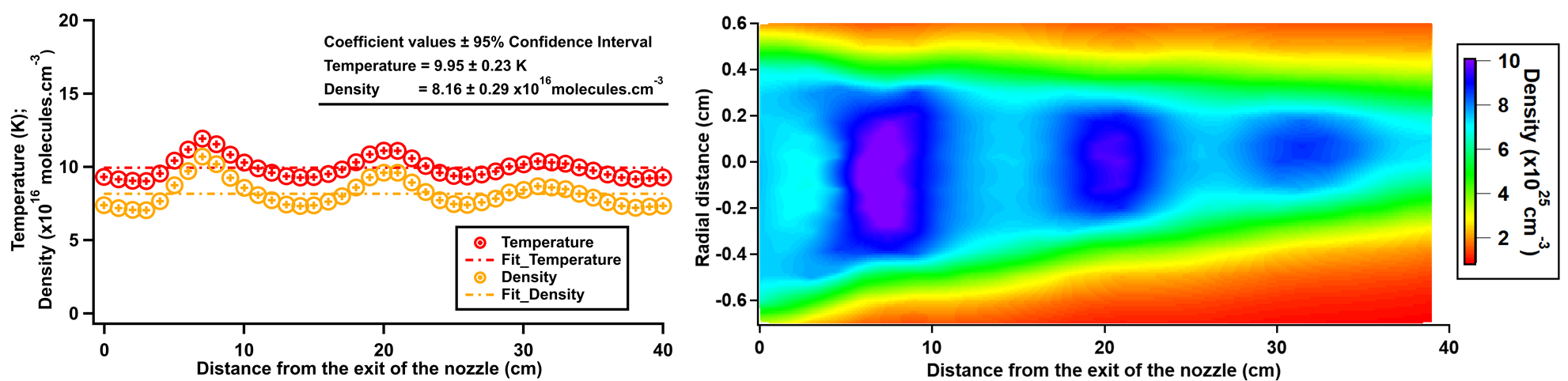}
    \caption{Pitot measurements for the He10K nozzle. The left panel shows the measurements along the central axis of the nozzle and right panel displays the 2D-mapping of the calculated density.}
    \label{fig:He10K}
\end{figure}

\begin{figure}
\centering
    \includegraphics[width=\linewidth]{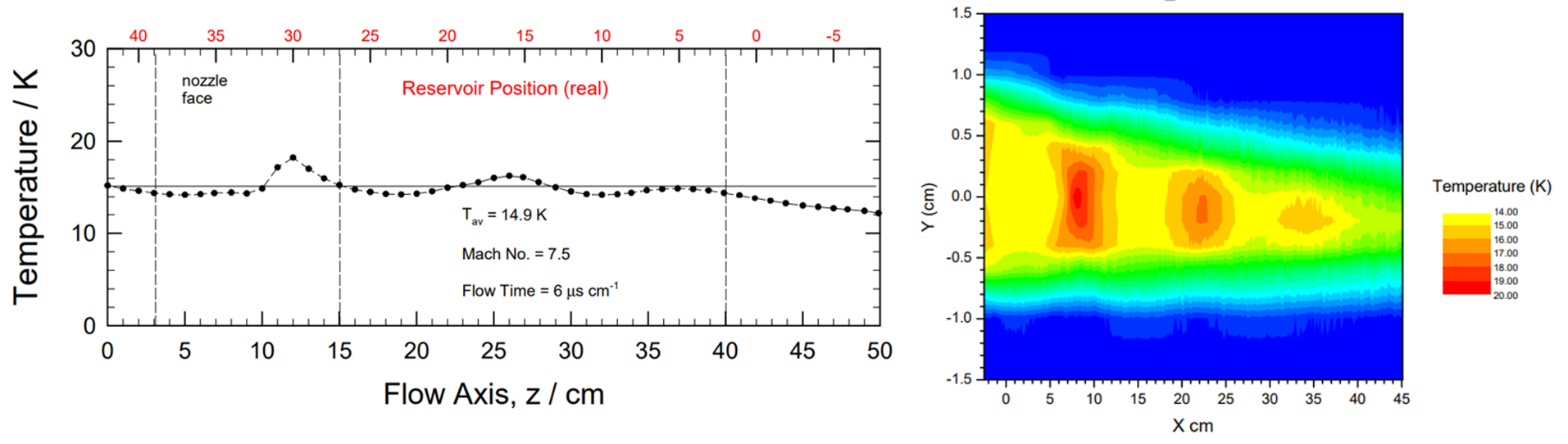}
   
    \caption{Pitot measurements for the He15K nozzle. The left panel shows the measurements along the central axis of the nozzle and right panel displays the 2D-mapping of the temperature. Note that the temperature outside the isentropic core is lower due to the invalidity of isentropic equations in boundary layers.}
    \label{fig:He15K}
\end{figure}

\begin{figure}
\centering
    \includegraphics[width=0.86\linewidth]{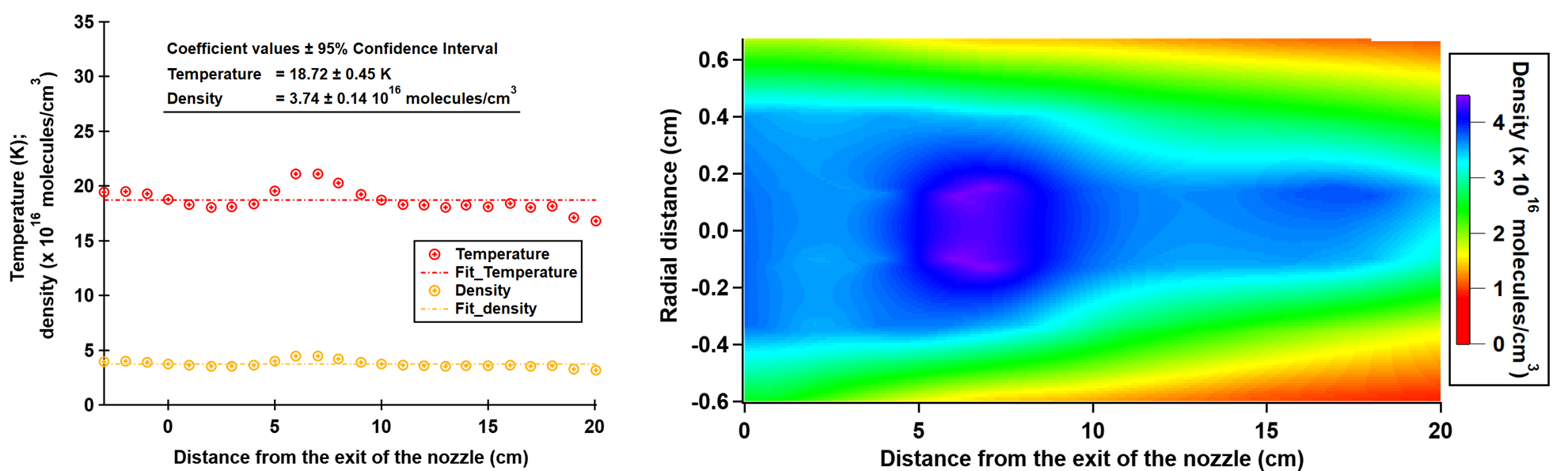}
     
    \caption{Pitot measurements for the He20K nozzle. The left panel shows the measurements along the central axis of the nozzle and right panel displays the 2D-mapping of the calculated density.}
    \label{fig:He20K}
\end{figure}

\begin{figure}
\centering
    \includegraphics[width=0.86\linewidth]{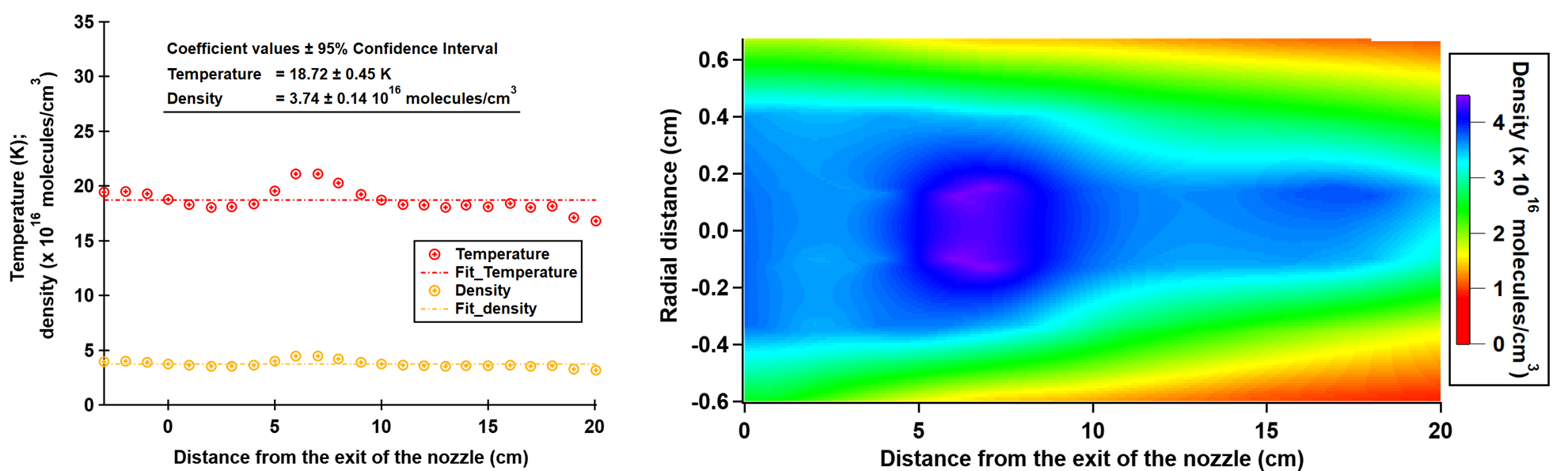}
   
    \caption{Pitot measurements for the He35K nozzle. The left panel shows the measurements along the central axis of the nozzle and right panel displays the 2D-mapping of the calculated density.}
    \label{fig:He35K}
\end{figure}

Uniform supersonic flows at different temperatures are obtained with different Laval nozzles and their Pitot (impact pressure) measurements are reported in Figs. \ref{fig:He10K},\ref{fig:He15K}, \ref{fig:He20K}, \ref{fig:He35K}.

\section{\label{AppC} Collision-energy dependence of the pressure-broadening cross-sections}

\begin{figure}[!ht]
    \centering
    \includegraphics[width=0.35\linewidth]{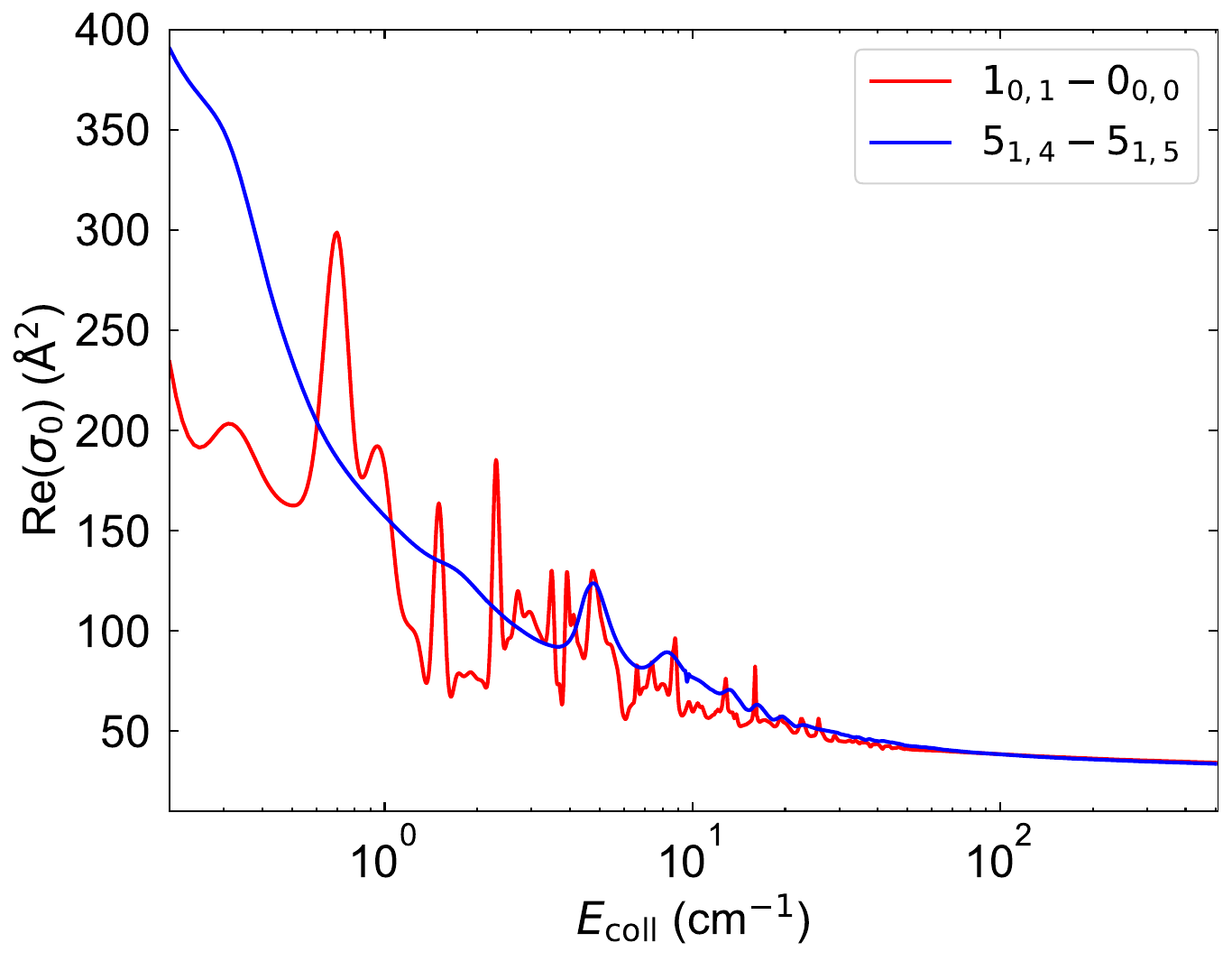}
    \hspace{0.03\linewidth}
    \includegraphics[width=0.36\linewidth]{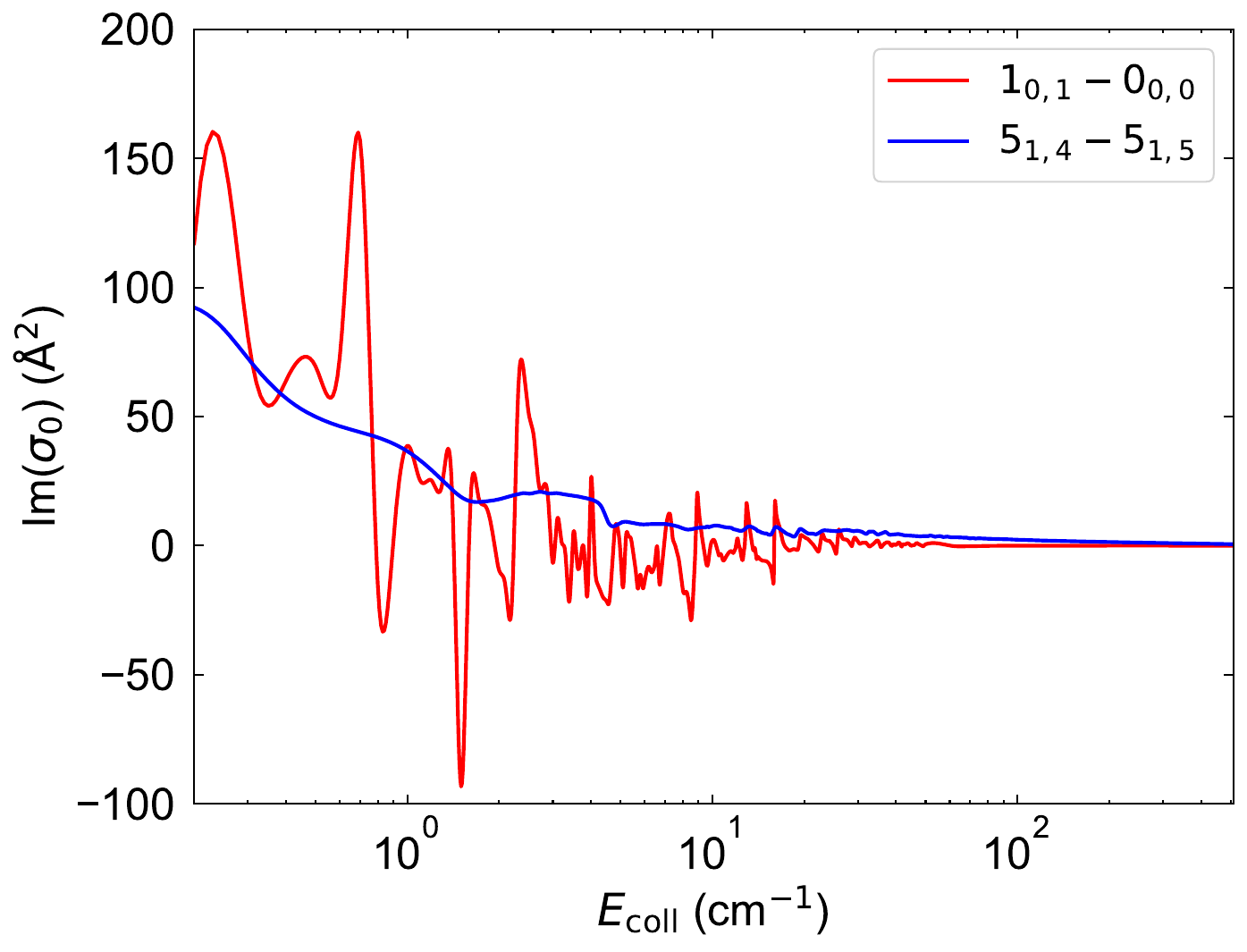}
    \caption{Pressure-broadening $\operatorname{Re}[\sigma^{(q)}_0]$ and pressure-shift $\operatorname{Im}[\sigma^{(q)}_0]$, cross-sections as a function of collision energy for the He-perturbed $1_{0,1}$--$0_{0,0}$ {\it para}-H$_2$CO and $5_{1,4}$--$5_{1,5}$ {\it ortho}-H$_2$CO transitions, calculated from the close coupling $S$-matrices.}
 \label{fig:XSplot}
 \end{figure}

Fig.~\ref{fig:XSplot} shows the collision-energy dependence of the generalized spectroscopic pressure-broadening cross-sections $\operatorname{Re}[\sigma^{(q)}_0]$ for the He-perturbed $1_{0,1}$--$0_{0,0}$ {\it para}-H$_2$CO and $5_{1,4}$--$5_{1,5}$ {\it ortho}-H$_2$CO transitions, as calculated from the close coupling $S$-matrices. As can be seen, the two transitions exhibit rather different qualitative behaviors, notably due to the strong resonance nature of the $0_{0,0}$--$1_{0,1}$ transition. On the other hand, quantitatively these cross-sections are rather similar, both show a monotonic decrease with energy. The most notable differences are at low collision energies (especially below $\sim 10$~cm$^{-1}$), which also explains the differences seen in the derived low-temperature cross-sections shown in Fig.~\ref{PB_xsec}. This might be explained by the fact that the lowest {\it para}-H$_2$CO states are coupled to only a few other higher states in this regime, while there are a much higher number of couplings in the case of the $5_{1,5}$ and $5_{1,4}$ {\it ortho}-H$_2$CO states, and therefore the resulting resonances strongly overlap in this case.
 
Fig.~\ref{fig:XSplot} also shows the pressure shift cross-sections, $\operatorname{Im}[\sigma^{(q)}_0]$, which are close to zero for the collision energies that contribute to the temperatures investigated. This shift is noticeably small relative to the pressure broadening width and hence is not observed in the current experiments.

\end{appendix}

\end{document}